\documentclass{lmcs}
\pdfoutput=1

\usepackage{lastpage}
\lmcsdoi{15}{2}{8}
\lmcsheading{}{\pageref{LastPage}}{}{}%
{Jan.~27,~2018}{Apr.~30,~2019}{}


\usepackage{graphicx}
\usepackage{url}
\usepackage{hyperref}

\usepackage{verbatim}
\usepackage{latexsym}
\usepackage{amssymb}
\usepackage{amsfonts}
\usepackage{amsmath}
\usepackage{amsbsy}
\usepackage{enumerate}
\usepackage{cite}

\DeclareFontFamily{OT1}{msbm}{}
\DeclareFontShape{OT1}{msbm}{b}{n}{<5> <6> <7> msbm7 <8> msbm8 <9> msbm9  <10> <10.95> <12> 
<13.82> <14.4> <16.59>  <17.28> <19.91> <20.74> <23.89>  <24.88> msbm10}{}
\DeclareMathAlphabet{\bb}{OT1}{msbm}{b}{n}

\DeclareFontShape{OT1}{cmr}{m}{it}{<5> cmr5 <6> cmr6 <7> cmr7 <8> cmr8 <9> cmr9  <10> cmr10 <10.95> <12> cmr12 <13.82> <14.4> <16.59> <17> cmr17 <17.28> <19.91> <20.74> <23.89> <24.88> cmr17}{}

\DeclareFontShape{OT1}{cmtt}{m}{n}{<5> <6> <7> <8> cmtt8 <9> cmtt9  <10> <10.95> <12> 
<13.82> <14.4> <16.59>  <17.28> <19.91> <20.74> <23.89> <24.88> cmtt10}{}

\catcode`@=11 

\def\barz #1,#2{#1_0,\ldots,#1_{#2}}     
\def\baro #1,#2{#1_1,\ldots,#1_{#2}}     
\def\barj #1,#2,#3{#1_{#2},\ldots,#1_{#3}}  

\def\header#1{\bigskip\par\noindent {\bf #1}\quad}

\def\factref#1{({\bf #1})}

\def\bpf#1{\relax\header{Proof \if !#1\else(#1)\fi:}}

\def\sqr#1#2{{\vcenter{\vbox{\hrule height.#2pt
      \hbox{\vrule width.#2pt height#1pt \kern#1pt \vrule width.#2pt}
      \hrule height.#2pt}}}}%

\def\mfor{\quad\hbox{for}\quad}

\def\dalign#1{\vtop{\openup-\jot\ialign{&$##$\hfil\cr#1}}}

\def\ldisplayindent{\quad}
\def\ldisplaylines#1{\displ@y\openup\jot
  \halign{\hbox to\displaywidth{$\ldisplayindent\displaystyle##\hfil$}\crcr
    #1\crcr}}

\def\ldisplaytwo#1{\displ@y\openup\jot
  \halign to\displaywidth{\rm ##\quad \tabskip\z@skip 
       &$\displaystyle{{}##}$\hfil\tabskip=0pt plus 1000pt minus 1000pt\crcr
    #1\crcr}}

\def\ldisplaythree#1{\displ@y\openup\jot
  \halign to\displaywidth{\rm ##\quad \tabskip\z@skip 
       &$\displaystyle{{}##}$\hfil\tabskip=0pt plus 1000pt minus 1000pt\crcr
       &$\displaystyle{{}##}$\hfil\tabskip=0pt plus 1000pt minus 1000pt\crcr
    #1\crcr}}

\def\hcr{\hidewidth\cr}

\def\specfootnote#1{
      \insert\footins\bgroup\notesize
      \interlinepenalty100 \let\par=\endgraf
        \leftskip=\z@skip \rightskip=\z@skip
        \splittopskip=10pt plus 1pt minus 1pt \floatingpenalty=20000
        \smallskip
        \noindent\hangindent 5pt \hbox to 5pt{\hss $^{#1}$}
        \bgroup\strut\aftergroup\@foot\let\next}

\catcode`@=12 

\def\tfun{\rightarrow}
\def\pfun{\buildrel {\scriptscriptstyle\rm p}\over \rightarrow}

\def\bnd{\mathrel{:=}}
\def\subst#1#2{\{#1\bnd #2\}}

\def\ls{{ \{ }}
\def\rs{{ \} }}

\def\ldbk{[\![}
\def\rdbk{]\!]}
\def\cx#1{\ldbk\if !#1\ \else{#1}\fi\rdbk}
\def\hol{\varepsilon}

\def\limp{\,\Rightarrow\,}
\def\liff{\>\Leftrightarrow\>}

\def\land{\,\wedge\,}
\def\setbar{\;\mathrel{\strut\vrule}\;}

\def\betav{\beta_{\rm v}}

\def\bnd{\mathrel{:=}}

\def\subst#1#2{{\ls #1\bnd #2\rs}}

\def\mimp{\quad\hbox{\rm implies}\quad}

\def\range{\mathrel{\hbox{ranges over}}}

\def\mabbreviates{\quad\hbox{\rm abbreviates}\quad}

\catcode`@=11 

\def\ldisplaythree#1{\displ@y\openup\jot
  \halign to\displaywidth{\rm ##\quad \tabskip\z@skip
       &$\displaystyle{{}##}$\hfil\tabskip=0pt plus 1000pt minus 1000pt
       \quad
       &$\displaystyle{{}##}$\hfil\tabskip=0pt plus 1000pt minus 1000pt\crcr
    #1\crcr}}

\catcode`@=12 

\def\ldisplayindent{\quad}

\def\lsbk{[}
\def\rsbk{]}

\def\setbar{\,\mathrel{\strut\vrule}\,}

\def\setbar{\,\mathrel{\strut\vrule}\,}

\def\red{\buildrel \ast \over\mapsto}

\def\ired{\mapsto}
\def\opeq{\cong}

\def\Isdef{\mathop{\downarrow}}

\def\Mcx{{\bb M}}

\def\Rcx{{\bb R}}

\def\hol{\bullet}

\def\cx#1{\lsbk\if !#1\ \else{#1}\fi\rsbk}
\def\pcx#1{\ldbk\if !#1\ \else{#1}\fi\rdbk}

\def\rcx{{\it R}}
\def\mcx{\Gamma}

\def\vexp{{\it v}}
\def\exp{{\mathnormal e}}

\def\vsub{\sigma}

\def\qapp{{\tt app}}
\def\qlet{{\tt let}}
\def\qseq{{\tt seq}}
\def\qfix{{\tt Y}}
\def\qeq{{\tt eq}}

\def\qt{{\tt t}}
\def\qnil{{\tt nil}}

\def\qif{{\tt if}}

\def\qcell{{\tt cell?}}
\def\qmk{{\tt mk}}
\def\qset{{\tt set}}
\def\qget{{\tt get}}

\def\qpair{{\tt pair}}

\def\ucx{{\it U}}

\def\fmla{\Phi}

\def\qz-cell{{\fmla_{\hbox{\rm 0-cell}}}}
\def\qu-cell{{\fmla_{\hbox{\rm 1-cell}}}}
\def\qnowrite{{\fmla_{\lnot{\rm write}}}}

\def\ssqr#1#2#3{{\vcenter{\vbox{\hrule height.5pt
      \hbox{\vrule width.5pt height#1pt \kern#2pt{#3}\kern#2pt\vrule width.5pt}
      \hrule height.5pt}}}}%

\def\qnonexp{{\fmla_{\lnot{\rm expand}}}}

\def\Val{{\bf Val}}
\def\Nil{{\bf Nil}}
\def\Cel{{\bf Cell}}

\def\cNat{{\bf Nat}}
\def\mfun{\,{\buildrel \mu\over \rightarrow}\,}

\def\tfun{\rightarrow}

\long\def\omitthis#1{\relax}

\def\blangle{\Big\langle\mskip-8mu\Big\langle}
\def\brangle{\Big\rangle\mskip-8mu\Big\rangle}
\def\mkc#1#2#3#4{ \blangle\, {#2} \setbar {#3}\,\brangle_{#4}^{#1}}

\def\qletactor{{\tt letactor}}
\def\qbecome{{\tt become}}
\def\qsend{{\tt send}}

\def\qevent{{\tt event}}
\def\qtick{{\rm tick}}
\def\beh{{b}}

\def\Ticker{{\rm Ticker}}

\def\striso{\simeq}

\begin{document}

\title{
Reasoning about effects: from lists to cyber-physical agents
}

\author[I.A.~Mason]{Ian A. Mason}
\address{SRI International,Menlo Park, CA 94025, USA}
\email{\{ian.mason,carolyn.talcott\}@sri.com}

\author[C.Talcott]{Carolyn Talcott}

\maketitle

\begin{abstract}

Theories for reasoning about programs with effects initially
focused on basic manipulation of lists and other mutable
data. The next challenge was to consider higher-order
programming, adding functions as first class objects to
mutable data. Reasoning about actors added the challenge of
dealing with distributed open systems of entities interacting
asynchronously. The advent of cyber-physical agents introduces
the need to consider uncertainty, faults, physical as well as
logical effects. In addition cyber-physical agents have
sensors and actuators giving rise to a much richer class of
effects with broader scope: think of self-driving cars,
autonomous drones, or smart medical devices.

This paper gives a retrospective on reasoning about effects
highlighting key principles and techniques and closing with
challenges for future work.

\end{abstract}

\omitthis{
Dedication TO APPEAR
}

\section{Introduction}\label{intro} 

``Real programs have effects--creating new structures, examining and
modifying existing structures, altering flow of control, etc.'' This
was the first sentence in our 1991 paper published in the debut of the
Journal of Functional Programming \cite{mason-talcott-91fp}. According
to the Oxford dictionary an \emph{effect} is ``a
change that is a result or consequence of an action or other cause.''
In the computational world, effects can be broadly characterized as
Read, Write, or Allocation/Creation effects. Examples include mutable
data, objects with local state and methods for access, and actors. The
effect, in terms of semantic foundations and reasoning principles, of
allowing effects depends on what other capabilities a language or
computational model provides, for example: first-order vs
higher-order, sequential vs concurrent/distributed.

Fast forward to the present, and the sentence has a much broader
meaning in which effects include interacting with and acting on the
external environment: self-driving cars and aircraft, medical devices,
automated manufacturing, automated biology experiments, smart homes,
\ldots.

Equivalence between data structures or active entities is a key
concept to be addressed in any system for reasoning about programs.
The good news is that 
the rule ``replacing equals by equals gives equals'' is
usually achievable for a suitable notion of equivalence. The other
property of equality that is important in many logics is that
replacing a variable by some expression preserve equality.  This
fails when evaluation of expressions has effects.
What about other laws of equivalence in the presence of effects? What
other properties do we want to reason about in general (as opposed to
application specific properties)? What about types vs sets defined by
a property? What are some helpful reasoning principles or proof
schemes?

The work presented here builds on three main themes, corresponding to
works that have guided our approach.
First, the languages we consider share key features of what we call
Landinesque languages in the spirit of Landin's seminal papers
\cite{landin-64cj,landin-66cacm}.  Such languages have a functional
core extended by primitives for data and control operations and
coupled with an operational semantics structured to support 
modular extension and equational reasoning.  
Second, satisfying laws of the computational lambda calculus
\cite{moggi-88complam,moggi-89} by the functional core is a key requirements for notions of equivalence. 
Third, our approach to reasoning and logical formalization is based on
Feferman's methods for formalization of constructive mathematics
and his ideas concerning variable types~\cite{feferman-75explicit,feferman-90poly}.
Operational equivalence, being indistinguishable by any enclosing program,
has generally been an important approach to defining equivalence of expressions, starting with Plotkin's work~\cite{plotkin-75tcs}. 
The notion of uniform semantics provides an important tool
for deriving laws of operational equivalence, avoiding the need
to explicitly reason about ``all enclosing programs''.
In particular, a uniform semantics allows one to compute symbolically
with contexts and delay instantiation of variables until they are used.

\omitthis{
Operational equivalence, being indistinguishable by any enclosing program,
has been an important approach to defining equivalence of expressions,
but as systems become more complex it begins to show some wear and tear.
We take a look at this and the above issues in increasingly complex systems.
}

In the \S 2 we look at the simplest example of effects: mutable data
in a first order language. This was also historically where our study
of effects began. We then, in \S 3,  move to the richer world of higher order programming
in the presence of mutable data. In both these cases the world is sequential and deterministic.
In \S 4 we look at distributed systems, where the notions of sequentiality, determinism, and
even termination no longer play center stage. In \S 5 we touch on the newer cyber-physical world,
and the issues that arise therein. Finally in \S 6 we summarize
and make some concluding remarks on the challenges
we have uncovered.

\omitthis{
notions of defined  -- not same as is a value

Such facilities are important not only for optimization but also for
communication, clarity and simplicity in programming. Thus it is
important to be able to reason both informally and formally about
programs with effects and not to sweep effects either to the side or
under the store parameter rug.
}

\omitthis{
General issues
  approximation/equivalence
  replacability
  classifications and other properties
  
What goes wrong
Context lemmas?
}


\section{First Order Theory of Mutable Data}\label{folists}

\omitthis{
S-Expression -- canonical mutable data structures
  basics of effects
  sample equational laws  
  what goes wrong
}

The simplest examples of effects are those usually slandered as side
effects: variable assignment and mutable data such as pointers,
arrays, and lisp style lists (or cons cells).  Initially we
concentrated on the Lisp cons cells, but eventually became more
enamored with the ML style reference.
$\qmk$ is a memory allocation primitive: the evaluation of $\qmk(v)$
results in the allocation of a {\bf new} 
{\em memory cell} and initializes this cell so that it contains 
the value $v$. The value returned by this call to $\qmk$ is the newly
allocated cell. $\qmk$ is total.
$\qget$ is the memory access primitive: the evaluation of $\qget(v)$
is defined iff $v$ is a memory cell. If $v$ is a memory cell,  then 
$\qget(v)$ returns the value stored in that cell. Note that there is 
no reason why a cell cannot store itself (or some more elaborate cycle).
$\qget$ is partial.
$\qset$ is the memory modification primitive: 
the evaluation of $\qset(v_0,v_1)$ is defined iff 
$v_0$ is a memory cell. If $v_0$ is a memory cell,  
then $\qset(v_0,v_1)$ modifies that cell so that its new 
contents becomes $v_1$. The value returned by a call to $\qset$ is somewhat
arbitrary and somewhat irrelevant. We have chosen $\qnil$ as
the return value,
thus if $v$ is a cell, then 
$\qset(v,v)$ will return $\qnil$, and more importantly modify 
$v$ so that it contains itself. $\qset$ is partial.

Mutable data structures are richer than immutable ones since the
ability to mutate allows one to distinguish between objects that have
identical structure, but are not the same object stored in memory. The
notion of being the same object in memory, in the Lisp tradition, is
known as $\qeq$-ness, or being $\qeq$ rather than just equal.

We illustrate this phenomena by providing a function that returns \verb|t| if the reference cells x and y are
the same object in memory, and \verb|nil| otherwise.
$$\ldisplaylines{
\dalign{
\lambda x . \lambda y . \qlet&\ls x_0 \bnd \qget(x), y_0\bnd \qget(y)\rs\hcr
		             &\qseq(&\qset(x,\qnil),\cr
			     &      &\qset(y, \qt),\cr
			     &      &\qlet\ls z \bnd \qget(x)\rs\qseq(\qset(x, x_0), \qset(y, y_0), z))\cr}
\cr}$$
      
It is important to notice that the function above leaves the state of
memory completely unchanged, even though during execution, observable
modifications are made. As a result the function would be
indistinguishable from the pure version which relies on the eq
primitive found in Lisp languages
$$\ldisplaylines{
\lambda x . \lambda y . \qeq(x, y)
\cr}$$
\noindent assuming a simple (single threaded) notion of indistinguishablity. We
can make this observation more formal by using the notion of a
context, an expression with a hole, $\hol$, or more pragmatically an enclosing
program.  We say two expressions, $f$ and $g$, are operationally
equivalent iff $C\cx{f}$ gives the same result computationally as  $C\cx{g}$
for any closing context C in the language at hand. The notion of sameness can
usually be taken to be a very coarse grained notion such as simply
being defined.

We introduce contexts at this early stage because they turn out to be
crucial in the study of languages with effects. They can be used to
define the semantics of programs by elegant reduction systems. As we
have already seen they can be used to define the notion of
computational indistinguishablity, and they can even be used as a
logical construct to express properties of programs, akin to a Hoare triple.
A contextual assertion takes the form,  
$\ucx\pcx{\Phi}$, and asserts that the assertion $\Phi$ holds at the point in the computation $\ucx$
when the $\hol$ is reached.
A simple example of this 
is the axiom which expresses the \emph{allocation} effects of $\qmk$:
$$\ldisplaylines{
 \qlet\ls x\bnd\qmk(v)\rs
             \pcx{\lnot(x\opeq y) \land \qcell(x)\opeq \qt\land 
               \qget(x)\opeq v}
             \cr}$$
Intuitively it asserts that the result of a call to $\qmk(v)$ is a cell
whose contents is $v$ and more importantly, different from every value that
existed prior to the call. Contextual assertions are first class formulas and
can be quantifed, and be passed to the boolean connectives. Thus we can make the implicitly
universally quantified value $y$ explicitly quantified:
$$\ldisplaylines{
(\forall y)(\qlet\ls x\bnd\qmk(v)\rs
             \pcx{\lnot(x\opeq y) \land \qcell(x)\opeq \qt\land 
               \qget(x)\opeq v})
             \cr}$$
This also is a good illustration of the fact that we make no distinction between
logical variables, and the variables of our programming language. They are one and the same.
We will discuss contextual assertions in more detail in section \ref{vtloe}.

In modeling first order languages with mutable data one must
have some representation of the current state of the the
data structures at hand. In first order Lisp like languages
the state of memory can be simply represented by a \emph{memory context}, an
expression, or context, of the form
$$\ldisplaylines{
\qlet\{z_1 \bnd \qmk(\qnil)\}
\ldots
\qlet\{z_n \bnd \qmk(\qnil)\}
\qseq(
    \qset(z_1,v_1),
      \ldots,
    \qset(z_n,v_n),
    \hol)
\cr
}$$
The set of memory contexts, $\Mcx$, is the set of contexts $\mcx$ 
of the above form 
where $z_i\not= z_j$ when $i\not= j$.
Subsequently $\mcx\range\Mcx$. Here we have used unary cells, the definition
for binary cells is entirely analogous.
Note that we split the construction of memory into allocation followed by assignment
to allow for the construction of arbitrary, possibly cyclic, memory.
That memory can be represented as syntactic
contexts simplifies the expression of many properties since it
provides natural notions of parameterized memory objects, of binding, and
of substitution for parameters.  We define a reduction calculus on syntactic entities,
$$\mcx_0; e_0 \ired \mcx_1; e_1$$ called descriptions. They consist of a memory context, a syntactic representation of
the state of memory, and an expression, representing the computation taking place.
The current computation can be further divided into the current instruction, and the current continuation. Their syntactic counterparts are 
{\it redexes},   and {\it reduction contexts}, respectively.
Redexes describe the primitive computation steps.  A primitive step is either
a $\betav$-reduction
or the application of a primitive operation to a
sequence of value expressions.
Reduction contexts,  $\Rcx$, identify the subexpression of an expression
that is to be evaluated next, they 
correspond to the standard reduction strategy
(left-first, call-by-value)
of~\cite{plotkin-75tcs} and were first 
introduced in~\cite{felleisen-friedman-86fdpc}.
We use $\rcx$ to range over $\Rcx$.

In addition, the syntactic representation of computation state 
allows us to compute with open expressions and provides a natural
scoping mechanism for memory simply using laws for bound variables. Many
of the basic equivalence relations on memories and other semantic
entities translate naturally into simple syntactic equivalences such as
alpha equivalence.

Reasoning about programs with effects is more delicate than the pure
or effect-free languages. For example, it is not the case that substitution
instances of equivalent expressions are equivalent
$\qeq(x, x)$
will always evaluate to \verb|t| in a world of atomic data, references, and cons cells,
but the substitution instance $\qeq(\qmk(x), \qmk(x))$
will always be false. This is simply because the evaluation of an expression can have effects,
and evaluating an expression more than once can be noticeable. 
Again, one is rescued by
contexts, since the property that remains true can be captured by
$$\ldisplaylines{
\qlet\{x \bnd e\}\qeq(x, x)
\cr}$$
always evaluating to \verb|t|, hinting at the crucial role contexts can make in being able to express
subtle properties of the primitives involved.

The quintessential property of operational equivalence is that it is a
congruence relation.  $e_0 \opeq e_1$ implies $C\cx{e_0} \opeq
C\cx{e_1}$ for any context $C$, making it an ideal tool for reasoning
symbolically about programs with effects. The down side to operational
equivalence is that it is in general very hard to establish
equivalences.  In the case of first order lisp programs this
difficulty is surmounted by defining a seemingly stronger perspicuous
relation called \emph{strong isomorphism}, and establishing that it
implies operational equivalence.

In \cite{mason-csli,mason-scp} \emph{strong isomorphism} is defined
between two expressions $e_0$ and $e_1$, written $e_0 \striso e_1$, if
and only if for every closed instantiation\footnote{a closed
instantiation is a substitution of values for the free variables that
results in a closed expression, where the notion of closed maybe
relative to the memory context at hand.} the expressions evaluate to
equal values in states that are identical, modulo the production of
garbage. Here garbage is used to describe memory that is not reachable
from either the result, or the original memory.  Simple examples of
strongly isomorphic expressions are
$$\ldisplaylines{
  \qeq(x,x)\striso \qt\cr
  \qseq(\qset(x, v), \qset(x, w)) \striso \qset(x, w)\cr
  \qseq(\qmk(x), \qmk(u)) \striso \qmk(u)
\cr}$$
the first two simply evaluate to identical states, the third does so too,
but produces some garbage along the way.
The main result concerning \emph{strong isomorphism}, apart from its usefulness
in establishing equivalences, is that in the first order Lisp world it coincides
with operational equivalence, and so can be used as a tool to establish the operational equivalence
of expressions.

In \cite{mason-talcott-89completeness,mason-talcott-93tcs} we used
this characterization and the ability to reason syntactically to provide a
formal system for establishing operational equivalence of first order Lisp
like programs, and showed that it was sound. The system was also shown
to be complete when restricted to non-recursive programs.

Note that \emph{sequentiality} is very important in establishing the above results.
In a multi-threaded world strong isomorphism would not coincide with operational
equivalence, since multi-threaded contexts would be sensitive not just to
the result of the computation, but also to the state of the world at every step.
Without some form of mutual exclusion one would not be able to define $\qeq$-ness
in terms of mutation, since a process running concurrently could also be mutating the cells being tested.


\section{Reasoning about Functions and Effects}\label{vtloe}

Treating functions as first class entities, with the ability to create
functions during execution, and to store or return functions as values
adds new complications for reasoning about programs.
To begin with, equality of values in the usual sense is
no longer decidable.  This of course is an issue even in
the absence of mutable data.  Another feature is the 
ability to create functions that share one or more 
instances of mutable data structures 
For example 
$$\ldisplaylines{
\qlet\{ x \bnd \qmk(\qnil)\}\qlet\{ f \bnd \lambda y . e_0(x)\}\qlet\{ g \bnd \lambda z . e_1(x)\}\qpair(f, g)
\cr}$$
constructs two closures, $f$ and $g$, that share a reference cell $x$.
Thus $f$ and $g$ can communicate with each other via setting the value stored in $x$
or export $x$ for other expressions to manipulate. Properties of $f$ and $g$
may crucially depend on how the visibility of this $x$ is maintained,
making reasoning a  challenge~\cite{meyer-sieber-88popl,mason-talcott-92lics}.
A concrete example of this is a version of the call-by-value fixed point combinator,
let $\qfix_v$ be 
$$\ldisplaylines{
\dalign{\lambda y .\qlet&\{z \bnd \qmk(\qnil)\}\hcr
                       &\qseq(&\qset(z, \lambda x.
                                          \qapp(\qapp(y, \qget(z)),x)),\cr
                       &      &\qget(z))\cr}
\cr}$$
This version of the fixed-point combinator is essentially identical to the 
one suggested by Landin~\cite{landin-64cj}.  
When applied to a functional $F$ of the form $\lambda f. \lambda x.e$,
$\qfix_v$ creates a private local cell, $z$, with contents
$G =  \lambda x.\qapp(\qapp(F,\qget(z)),x)$, and returns $G$.
By privacy of $z$, $G$ is operationally equivalent to $F(G)$ 
(cf.~~\cite{mason-talcott-91fp}). 
Note that this example is typable in the simply typed lambda calculus
(for provably non-empty types (cf. ~~\cite{honsell-mason-smith-talcott-93ic})).
Thus adding operations for manipulating
references to the simply typed lambda calculus causes the failure of 
strong normalization as well as many other of its
nice mathematical properties. 

As another example, the usual notion of function satisfies the property
that each time it is applied to a given argument, the result is the same.
This is not the case when functions have memory!  Here is a function
that returns a different number each time it is called.
$$\ldisplaylines{
\qlet\{ x\bnd \qmk(0) \} \lambda y . \qlet\{ z \bnd \qget(x)\rs \qseq(\qset(x, z + 1), z)
  \cr}$$
Though simple, such an example can easily be elaborated, using the sieve
of Eratosthenes, to enumerate the prime numbers.

\subsection{Equivalence}

Early work on reasoning about equality in higher-order languages
includes Plotkin's work defining operational approximations and
equivalence for various lambda calculi \cite{plotkin-75tcs}, Felleisen's
(and students) work on reduction calculi for languages with effects
\cite{felleisen-thesis,felleisen-etal-87tcs}, 
and Moggi's work on computational monads for
a variety of computational primitives \cite{moggi-91ic}.
Moggi's equational laws of computational lambda calculus \cite{moggi-89}
are the core equational theory for lambda-based computational languages.

In \cite{mason-talcott-91fp} we developed a theory of operational
approximation and equivalence for a language that combines
(call-by-value) lambda calculus and Lisp-like mutable lists. Our
definition of operational equivalence extends the extensional
equivalence relations defined by Morris and Plotkin to computation over
memory structures. Equational laws and methods for proving equivalence
were developed building on \cite{mason-thesis,mason-talcott-93tcs}. This
work provided the foundation for a Variable Type Logic of Effects
\cite{honsell-mason-smith-talcott-93ic} which extended equational
reasoning with language for defining sets (properties) and principles
for reasoning about set membership.

Just as in \S~\ref{folists}, the basis of our definition of operational equivalence is a
small step operational semantics, defined using memory contexts to
represent memory state and reduction contexts to represent the
continuation of a computation and reduction rules to define the small
steps of a computation.
Two expressions $e_0$, $e_1$  are operationally equivalent
if for any closing context $C$, $C\cx{e_0}$ and $C\cx{e_1}$  
are equi-defined.  This looks identical to the definition
in the first-order case.  The difference is in the set of
possible contexts.
It is easy to see that this is a congruence relation so substitution
of equals for equals gives equals.  But, substitution of an 
expression into equals does not give equals.  The counter-example
shown in \S~\ref{folists} remains a counter-example.

The definition of strong isomorphism in the first order case can be
lifted to our higher-language in an entirely analogous fashion
and just as in the first-order case, we have that 
\begin{itemize}
  \item strong isomorphism implies operational equivalence.
\end{itemize}  
\noindent
A key feature of $\striso$ is that reduction rules of the operational
semantics are a subset of the $\striso$ relation.   Thus many laws can be
proved by showing two expressions have a common reduct.
For example
$$\ldisplaylines{
\qlet\{z\bnd \qmk(x)\}\qseq(\qset(z,w),e) \striso \qlet\{z\bnd \qmk(w)\}e
\cr}$$
\noindent if $z$ and $w$ are distinct variables.
Furthermore, many of the laws of strong isomorphism from the first-order case 
continue to hold as laws of operational equivalence in the higher-order case, including laws based on reductions that do not directly involve functions.

The $\striso$ laws combined with congruence entail that the $\eta$ 
law of the lambda calculus holds in the sense that 
if $e$ denotes a function, i.e. $e \opeq \lambda x.e_0$, then
$e \opeq \lambda x.e(x)$.
In contrast, if we view the notion of function in the more general sense of being a lambda with local memory, 
$e \opeq \mcx\cx{\lambda x.e_0}$, then the $\eta$ law fails.
That is, in general $\lambda x.(\mcx\cx{\lambda x.e})x$ is
not operationally equivalent to $\mcx\cx{\lambda x.e}$.
As a counter-example take $\mcx$ to be $\qlet\{z\bnd\qmk(0)\}\hol$ and $\lambda x.e$ to be
$\lambda x.\qlet\{y\bnd \qget(z)\}\qseq(\qset(z,x),y)$.
Since in this case $\lambda x.(\mcx\cx{\lambda x.e})x$ is operationally equivalent to
$\lambda x . 0$, albeit with a memory leak, while $\mcx\cx{\lambda x.e}$ is a \emph{thunk} that when applied returns
the value it was previously applied to.

So how does the presence of higher-order entities distinguish 
$\opeq$ vs $\striso$?
Take any two distinct operationally equivalent lambda expressions, the simplest
pair that comes to mind is: $\lambda x . x$ and $\lambda x . \qseq(\qmk(0), x)$,
these are operationally equivalent, but not strongly isomorphic because, as values,
to be strongly isomorphic, they would have to be identical.

\omitthis{
Consider the expressions $e_0 = eq(x,x)$ and $e_1 = true$.
$e_0  \striso e_1$ in the first-order language, but not in
the presence of lambdas, because $eq(\lambda x.x,\lambda x.x) = false$.
Equality between functions is defined to be false as a
convenience, the intent being that the test is not allowed.  
Counterexamples can be constructed that do not rely on $eq$.
}

Strong isomorphism and computational reasoning based on reduction
rules nicely capture laws of local data and memory manipulation 
but there is much more to operational equivalence and
reasoning about \emph{all} contexts is daunting, even in the absence
of memory structures. 
Robin Milner's \emph{context lemma}~\cite{milner-77tcs}
showed that operational equivalence can be proved by considering
a small number of context patterns, thus greatly reducing the
complexity of proving operational equivalence laws.

An analog to the context lemma for languages with effects
is the  \emph{CIU (Closed Instantiations of Uses)} theorem 
which states that 
\begin{itemize}
\item
if all closed instantiations  of all uses of two expressions are
equidefined then the expressions are operationally equivalent.
\end{itemize}

A closed instantiation of a use of an expression $e$ is a closed
expression of the form $\mcx\cx{R[e^\sigma]}$  
where the memory context $\mcx$ and substitution
$\sigma$ represent the closed instantiation and the reduction context
$R$ represents the use.   
As hinted in the introduction, uniform semantics is key to
proving CIU.   Once established, CIU is used to develop methods for 
proving equivalence of  lambda functions with and without memory.

Using this theorem we can easily establish, for example, the validity of 
the Moggi's let-rules of the computational lambda calculus~\cite{moggi-89}
(see also~\cite{talcott-90disco} where these laws are established
for a language with control abstractions). 
$$\ldisplaytwo{
(i)&
 \qapp(\lambda x . \exp, \vexp) \opeq 
   \exp^\subst{x}{\vexp}\opeq \qlet\{x\bnd \vexp\}\exp
\cr
(ii)&
\rcx\cx{ \exp}\opeq \qlet\{x\bnd \exp\}\rcx\cx{x}
\cr
(iii)&
\rcx\cx{\qlet\{x\bnd \exp_0\}\exp_1}\opeq \qlet\{x\bnd \exp_0\}\rcx\cx{\exp_1}
\cr}$$
where in (ii) and (iii) we require $x$ not free in $\rcx$.

Another nice property that is easily established using CIU
is that reduction preserves operational equivalence:
$$\ldisplaylines{
\mcx;\exp \ired \mcx';\exp' \limp \mcx\cx{\exp} \opeq \mcx'\cx{\exp'}
\cr}$$
This property is the basis of the calculi found in~\cite{felleisen-hieb-92tcs}.
Our lambda language is an example of a \emph{Landinesque} language,
so called in the spirit of Landin's ``Next 700 Programming languages''
paper \cite{landin-66cacm}.
A key result is that the CIU theorem holds for any Landinesque language
with a suitably nice (uniform) semantics. Uniformity is captured by the ability to
compute with contexts rather than just expressions, and the exact notion
of uniformity is axiomatized in~\cite{mason-talcott-00feferman}.

\subsection{Formulas}

In addition to being a useful tool for establishing laws of operational
equivalence, CIU can be used to define a satisfaction
relation between memory contexts and equivalence assertions.
In an obvious analogy with the usual first-order Tarskian definition of 
satisfaction this can be extended to define a satisfaction relation 
$\mcx\models\fmla[\vsub]$ for formulas $\fmla$ and 
closing substitutions $\vsub$~\footnote{
Here a closing substitution binds at least the free variables not bound in 
the memory context. }.

The memory context $\mcx$ plays the role of the model, in that it specifies what objects
exist in memory, while the closing  substitution $\vsub$ binds variables to values that exist in
that model. Note that variables bound by the memory context $\mcx$ are cells, while
variables bound by the substitution $\vsub$ are arbitrary values.
The adjective closing just emphasizes that all free variables of $\mcx$
and $\fmla$ are in the domain of the substitution, and that no free variables creep in
amongst the values in the range of $\vsub$.

The atomic formulas of our language
assert the operational equivalence of two expressions.
In addition to the usual first-order formula constructions we add 
{\it contextual assertions:} if $\fmla$ is a formula
and $\ucx$ is a certain type of context, then $\ucx\pcx{\fmla}$ is a formula.
This form of formula  expresses the fact that
the assertion $\fmla$ holds at the point in
the program text marked by the hole in $\ucx$, if execution of the program
reaches that point.
The contexts allowed in contextual assertions are called
{\it univalent contexts}, ($\ucx$-contexts).
They are the largest natural class of contexts 
whose symbolic evaluation is unproblematic. The key restriction is that 
we forbid the hole to appear in the scope of a (non-$\qlet$) lambda, 
thus preventing the proliferation of holes.

One simple consequence of the definitions are the following three 
principles for reasoning about contextual assertions:
a general principle for introducing contextual
assertions (akin to the rule of necessitation in modal logic); 
a principle for propagating contextual assertions through 
equations; and a principle for composing contexts (or collapsing
nested contextual assertions).
$$\ldisplaytwo{
(i) & 
\models \fmla \mimp \models \ucx\pcx{\fmla}
\cr
(ii) &
\ucx\pcx{e_0 \opeq e_1} \limp \ucx\cx{e_0} \opeq \ucx\cx{e_1}
\cr
(iii) &
\ucx_0\pcx{\ucx_1\pcx{\fmla}}\liff(\ucx_0\cx{\ucx_1})\pcx{\fmla}
\cr}
$$
Also, as we have already mentioned in section \ref{folists}
one can naturally express properties such as the allocation effects
of $\qmk$:
$$\ldisplaylines{
 (\forall y)(\qlet\ls x\bnd\qmk(\vexp)\rs
             \pcx{\lnot(x\opeq y) \land \qcell(x)\opeq \qt\land 
                  \qget(x)\opeq\vexp})
\cr}$$

\subsection{Classes}

Using methods developed by Feferman~\cite{feferman-75explicit,feferman-90poly} 
and applied to lambda languages with control operators~\cite{talcott-90disco},
we extend our theory to include a general theory of classifications 
(classes for short). 
With the introduction of classes,  principles such as structural induction, 
as well as principles accounting for the effects of an expression can easily 
be expressed.
Classes serve as a starting point for studying semantic notions of type.
As will be seen, direct representation of type inference systems
can be problematic, and additional notions maybe required to provide
a formal semantics.  Even here classes are likely to play an important role.

Class terms are  either class variables, 
class constants, or comprehension terms, $\ls x \setbar \fmla\rs$.
We extend the set of formulas to include 
class membership and  quantification over class
variables.
We define (extensional) equality and subset relations on classes
in the usual manner.
$$\ldisplaylines{
K_0 \subseteq K_1 \mabbreviates (\forall x)(x\in K_0 \limp x\in K_1)
\cr
K_0 \equiv K_1 \mabbreviates 
    K_0 \subseteq K_1 \land K_1 \subseteq K_0 
\cr
}$$
A simple example of a class is the set of reference cells that
contain values in a specific set $K$:
$$\ldisplaylines{
\Cel = \ls x \setbar \qcell(x) \opeq \qt \rs
\cr
\Cel[K] = \ls x \setbar \qcell(x) \opeq \qt \land \qget(x)\in K \rs
\cr
}$$
We can also express a variety of function spaces, the
simplest are total, partial and  memory.\footnote{We use the standard notation of  $\bar{x}$ to denote a sequence
$x_0, \ldots, x_n$ of variables.}
$$\ldisplaylines{
\bar{X} \tfun Y = \ls f \setbar
(\forall \bar{x}\in \bar{X})(\exists y \in Y)
\qapp(f,\bar{x})\opeq y\rs
\cr
\bar{X}\pfun{Y} = \ls f \setbar
(\forall \bar{x}\in \bar{X})(\forall y)
       (\qapp(f,\bar{x})\opeq y \limp y\in Y)\rs
\cr
\bar{X} \mfun Y = \ls f \setbar
(\forall \bar{x}\in \bar{X})(\qlet\ls y \bnd \qapp(f,\bar{x})\rs\pcx{y\in Y})\rs
\cr
}$$
So for example, the reference operations can be given \emph{types}
by
$$\ldisplaytwo{
(mk)&
\lambda x . \qmk(x) \in (X\mfun\Cel[X])
\cr
(get)&
\lambda x .\qget(x) \in\Cel[X]\tfun X
\cr
(set)&
\lambda x . \lambda y . \qset(x, y)\in \Cel \tfun (\Val\mfun \Nil)
}$$
Class membership expresses a very restricted form of non-expansiveness,
allowing neither expansion of memory domain nor change in contents of
existing cells.
To illustrate some of the subtleties regarding class membership, 
and  notions of expansiveness, consider the
following expressions:
$$\ldisplaylines{
e_0 = \lambda x . \qmk(\qnil)
\cr
e_1 = \qlet\ls z \bnd \qmk(\qnil)\rs \lambda x . z
\cr
e_2 = \qseq(\qif(\qcell(y),\qset(y,\qnil),\qnil),  \lambda x . \qmk(\qnil))
\cr               
e_3 = \qseq(\qif(\qcell(y),\qset(y,\qnil),\qnil), 
               \qlet\ls z \bnd \qmk(\qnil)\rs \lambda x . z)
\cr}$$
Then each of these expressions evaluates to a memory function mapping
arbitrary values to cells containing $\qnil$.  But they differ in
the effects they have.
$e_0$ is a value (and as such neither expands nor modifies memory). 
$e_1$ is not a value  and is expansive (its evaluation enlarges the domain of memory) but does not
modify existing memory.  
$e_2$ may modify existing memory, but does not expand it.
$e_3$ is expansive, and it may modify existing memory.
These observations can be expressed in the theory as follows.
Let $T$ be $\Val\mfun \Cel[\Nil]$, and $\qnowrite[\Cel](e)$ be as defined below.
Then
$$\ldisplaylines{
e_0\in T \land e_0\in\Val
\cr
e_j \not\in \Val \mfor 1 \le j \le 3
\cr
\qlet\ls x \bnd e_j\rs\pcx{x\in T} \mfor 0 \le j \le 3
\cr
\qnowrite(e_j) \mfor 0\le j \le 1
\cr
\qnonexp(e_j) \mfor j \in \ls 0,2\rs
\cr
}$$
Let $\qnonexp(\exp)$ stand for 
the formula
$$\ldisplaylines{
(\forall X)(X \equiv \Cel \limp \qseq(\exp,\pcx{X \equiv \Cel}).
\cr}$$
Then $\qnonexp(\exp)$ says that execution of $\exp$ does non expand
the memory, although it might modify contents of existing cells.
$\qnowrite(\exp)$ is defined as:
$$\ldisplaylines{
\dalign{
(\exists X)(&X \equiv \Cel \land \cr
   &(\forall x\in X)(\forall z \in\Val))(\qget(x)\opeq z \limp \qseq(e, \pcx{\qget(x)\opeq z})))\cr}
\cr}$$

\subsection{Classes vs Types: the functional case}

In \cite{feferman-90poly} Feferman proposes an explanation of ML types
in the variable type framework.  This gives a natural semantics to
ML type expressions, but there are problems with polymorphism, even in the 
purely functional case.  
The collection of classes is much too rich to be considered a type system. 
One problem that arises is that fixed-point combinators can not be
uniformly typed over all classes.  This problem arises even in the
absence of memory~\cite{ssmith-thesis,talcott-90disco}.
Let  $\qfix_v$ by  any fixed-point combinator
(such that $f(\qfix_v(f)) = \qfix_v(f)$). Then it is not the case that  
$$\ldisplaylines{
 f \in (C \tfun C) \limp \qfix_v(f) \in C
\cr}$$
for all function classes $C$ ($C\subseteq A\pfun B$ for some classes $A,B$).
 
Define  $P$ to be the class of {\it strictly} partial maps from $\cNat$
to $\cNat$:
$$\ldisplaylines{
P = \ls g \in {\cNat}\pfun{\cNat} \setbar (\exists n \in \cNat)(\lnot\Isdef g(n))\rs
\cr}$$
Let 
$$\ldisplaylines{
 f = \lambda p.\lambda n.\qif(\qeq(n,0), n, p(n-1))
\cr}$$
Then we can prove 
$$\ldisplaytwo{
(1) & f \in P \tfun P 
\cr
(2) & \qfix_v(f) \in \cNat \tfun \cNat
\cr}$$
\factref{1} follows by simple properties of $\qif$, $\qeq$ and arithmetic
\factref{2} follows by induction on $\cNat$ 
using the fixed point property of $Y$.
Consequently, $\lnot(\qfix_v(f) \in P)$ 

\subsection{Classes vs Types: the imperative case}

The situation becomes more problematic
when references are added, even in the simply typed (or monomorphic)
case.
Na\"ive attempts to represent ML types as classes fails in sense that
ML inference rules are not valid.  The essential feature
of the ML type system, in addition to the inference rules, is
the preservation of types during the execution of well-typed programs,
not just of the text being executed, but also of the contents of any
cell in memory. This requirement is a strong form of subject reduction. 
One that does not
seem to be expressible using classes (quantifying over types, whatever they 
may be, seems problematic).
Our analysis indicates that ML types are therefore
more syntactic than semantic.

\omitthis{

Logic
  Contextual assertions key to expressiveness
  
  []Phi -- holds in all expansions
  [*] Phi -- holds in all reachable (updates and expansions) 
    definable,
 invisible set, visible garbage,

forall -- quantifies over values of current memory
forall_s -- quantifies of values of all expansions of current memory
    []forall
Contextual reasoning principles  

--------------------
Classes -- constants aon comprehension  -- closed under obseq
Defining function spaces

NOT obey type rules

R[M[e]] -> M;R[e]

M; R[e] -> M'; R[e']  => M; R'[e] -> M'; R'[e']

returns value vs is value

VTLOE
In recent years various systems for reasoning about properties of programs written in general programming languages have been proposed  most notably Hoare's logic  (Apt, 1981)  Dynamic logic (Harel, 1984)
Reynolds Specification Logic (Reynolds, 1982), Moggi's metalanguage for computational monads (Moggi, 1991) and Pitt's Evaluation Logic  (Pitts, 1990)

We seek simplicity of reasoning and expressivity of descriptions and justify our system on the basis of its power in deriving properties of one canonical semantics rather than on its general applicability  Finally we use classical logic  the systems of Moggi and Pitts are based on constructive and categorical Logic  Similarly  as Tennett p oints out  Reynolds has given examples showing that  Specification Logic must be intuitionistic  i e  the classical law for double negation forces models of variables and assignment to be trivial

In this pap er we intro duce a variable typ ed logic of e ects inspired by the variable type systems of Feferman.  These systems are two sorted theories of op erations and classes initially develop ed for the formalization of constructive mathematics (Feferman 1975,1979)
and later applied to the study of purely functional languages 
(Feferman   1985,1990). 
VTLoE  (Variable Type Logic of Effects)  is organized in two levels.
The  first level is the  first order theory of individuals built on assertions of equality  (operational equivalence)  and contextual assertions. The second level extends the logic to include classes and class membership.

The theory allows for the construction of inductively 
defined sets and derivation of the corresponding induction principles.
Classes can be used to express, inter alia,  the non expansiveness of terms  (Tofte 1990).  Other effects can also be represented within the system.
These include read / write effects  (Lucassen  1987, Lucassen and Gifford      1988,  Jouvelot and Gifford  1991) and various forms of interference  
(Reynolds 1978,1982)

contextual assertions allow axiomatizing effects such as allocation
expressions with hole not in scope of lambda

define modalities
  []Phi -- holds in all expansions of memory
  [*] Phi -- holds in all reachable (updates and expansions) memories
  
CAs expressive, but also cause some problems as it can observe what
seems like private store.

e0 \opeq e1 => (let{x:=e0}[[Phi]] \opeq let{x:=e1}[[Phi]])
fails 
e0 = lambda y.y
e1 = let{z:=mk(lambda y.y)}(lambda w.app(get(z),w))
Phi = (x \opeq lambda y.y)

weak quantifier allows to express `no cells', `one cell' in memory.

extensionality by quantifying over memory
rules for CAs

CAs commute w prop connectives 

if traps(U) << xbar
[*]((forall xbar) Phi) => [*](U[[Phi]])  if traps(U) << xbar

[*]((forall xbar) (Phi0 => Phi1)) => ([*](U[[Phi0]]) => [*](U[[Phi1]]))

weak quantifier  over values in given memeory
strong quantifier over values in all expansions of the given memory

Phi => let{x := mk(y)}[[Phi]]   holds for strong quantifiers, not for weak

---------------
Class variables range over sets of values closed under operational equivalence

K X {X | W|}   x in X
In classical logic and type theory the notion of function space is an
important structuring and reasoning tool.
With the examples .. above in mind.  We ask what are useful notions
of function space.

In the presence of classes `congruence' fails
  not(e0 \opeq e1 /\ let{x:=e0}[x in K] => let{x:=e1}[x in K])

Proving p belongs to class P vs  p has type T.

X1 ... Xn -mu-> Y = 
  { f | (forall x1 in X1 ... xn in Xn) (let{y:=app(f,x1...xn)})[[y in Y]]}

X1 ... Xn --> Y = 
  { f | (forall x1 in X1 ... xn in Xn)
         (exists y in Y)(app(f,x1...xn) \opeq y)}

(forall X)(mk in X -mu-> Cell[X])
(forall X)(get in Cell[X] -> X)

Phi_not(write)
Phi_not(read)
Phi_not(expand)

fixpoint operators on Phi[X] and associated induction principles

N\"aive attempts to represent ML types as classes fails in sense that ML
inference rules are not valid. Its seems that the essential feature of ML
type system, in addition to the inference rules, is what is known as subject
reduction, i.e. the preservation of types during the execution of well typed
programs. Our analysis indicates that ML types are therefore more syntactic
than semantic and provide complimentary tools for reasoning about programs.

}


\section{Actors: Open Systems of Interactive Agents}\label{actors} 

\omitthis{
\begin{verbatim}
Two levels:  actor language, actor components

observational equivalence vs maymust stuff

actor algebra

bigstep transform

equational rules

fairness

\end{verbatim}
}

An actor is a unit of concurrent/distributed interactive computation.
Each actor encapsulates state.  It can receive messages; which may
cause it to change state; it can send messages, to actors it knows about;
and it can create new actors.  Communication by message passing is 
reliable, and asynchronous with fair message delivery~\cite{hewitt-thesis,agha-thesis}.
We can describe actor behaviors 
using lambda expressions augmented with actor primitives 
($\qbecome$, $\qsend$ and $\qletactor$) analogous to
describing computation over memory
structures by adding memory effect primitives~\cite{agha-mason-smith-talcott-96jfp}.
{$\qsend$} is for sending messages;
$\qsend(a,v)$ creates a new message with receiver $a$ and 
contents $v$ and puts the message into the message delivery
system.
$\qletactor$ is for actor creation.
$\qletactor\ls x\bnd\beh\rs\exp$ 
creates an actor with initial behavior $\beh$, making the
new address the value of the variable $x$.  The expression $\exp$
is evaluated in the extended environment.
The variable $x$ is also bound in the expression $\beh$, thus allowing
an actor to refer to itself if so desired.
{$\qbecome$} is for changing behavior; $\qbecome(\beh)$ creates an
anonymous actor to carry out the rest of the current computation,
alters the behavior of the actor executing the $\qbecome$ to be
$\beh$, and frees that actor to accept another message.  This provides
additional parallelism.  The anonymous actor may send messages or
create new actors in the process of completing its computation, but
will never receive any messages as its address can never be known.

A consequence of the actor interaction model is unbounded non-determinism.
A classic example is the Ticker actor that maintains a counter,
sends itself \texttt{tick} messages to increment the counter, and
responds to requests from other actors by sending the current counter value.
$$\ldisplaylines{
  \beh_\Ticker = \dalign{
 \qfix_v(\lambda b . \lambda c . \lambda m . \qif(&m = tick,\cr
  &\qseq(\qsend(\tau, tick), \qbecome(\qapp(b, c + 1)))\cr
  &\qseq(\qsend(customer(m), c), \qbecome(\qapp(b, c)))))\cr}
  \cr
  \Ticker = \qletactor\{ \tau \bnd  \beh_\Ticker \} \qseq(\qsend(\tau, tick), \tau)
\cr}$$
We avoid going into the details of messages as data structures by using
$customer(m)$ to denote the sender of the message.
The 
$\Ticker$
has the property that (assuming it is sent an initial $\qtick$ message)
for any natural number $n$ there is a computation where a request
results in sending a number greater than $n$.  This is because,
although the request is guaranteed to be delivered and receive a response,
any number of $\qtick$s can be delivered before the request.

The operational semantics for actor systems is given by a transition
relation on \emph{actor configurations}.
A configuration
$$\mkc{\rho}{\alpha}{\mu}{\xi}$$
can be thought
of as representing a global snapshot of an actor system with respect
to some idealized observer~\cite{agha-thesis}.
It contains a collection of actors $\alpha$, messages $\mu$, external
actor names $\xi$ , and receptionist names $\rho$.  
The sets of
receptionists and external actors are the interface of an actor
configuration to its environment.  They specify which actors are
visible and which actor connections must be provided for the
configuration to function. Both the set of receptionists
and the set of external actors may grow as the configuration evolves.

Several semantics have been defined for actor configurations \cite{talcott-98models}
differing by treatment of ordering relations among send/receive events
and level of detail \cite{baker-hewitt-77cpp}. The basic operational semantics is the set of traces of fair
executions given by a reduction relation as for Landinesque languages. What is
different is the presence of interactions with the external world -- transitions for input of messages
from external (unseen) actors (\texttt{in(msg)}), and output of messages to these
external actors (\texttt{out(msg)}).   

Although we have never done so, actor computation is uniform enough
for it to be represented as a Landinesque language, what is lacking
is the development of a syntax rich enough to represent configurations.
However, a somewhat more crucial distinction is that unlike the sequential case, neither
the notion of reducing to a value, nor deterministic computation,
nor the notion of a computation terminating
are central concepts. Rather they are side lined to the more infinitary
notion of a computation path, and the collection of all computation paths.
It is in this infinitary realm that crucial questions of fairness arise
and play a part.  The unimportance of termination creates a new problem:
\emph{what are the primitive observations that underly any notion of
operational or observational equivalence?} The approach taken in~\cite{agha-mason-smith-talcott-96jfp}
is to introduce a primitive, $\qevent$, and observe
whether or not in a given computation, $\qevent$ is executed.  
This approach is similar in spirit
to that used in defining testing equivalence for 
CCS~\cite{denicola-hennessy-84tcs}, except that the required condition
of \emph{fairness} of actor computation simplifies matters by collapsing
two obvious candidates of equivalence into one. See \S 4 of \cite{agha-mason-smith-talcott-96jfp}
for more details.

With a notion of equivalence on actor expressions defined, a library
of useful equivalences can be established.
Since our reduction rules preserve the evaluation semantics of the
embedded functional language, many of the equational laws for 
the 
language of section~\ref{vtloe} (cf.~\cite{talcott-90disco-tcs})
continue to hold in the  actor language.
For example, the  laws of the untyped computational lambda
calculus~\cite{moggi-88complam}
continue to hold in the actor setting~\cite{agha-mason-smith-talcott-96jfp}.

Even though the actor language is not presented as a Landinesque language,
the fact that computation can be parametrically defined more generally on contexts
allows for laws to be established in an entirely analogous fashion to the CIU
principle. For example
if there is some $\exp'$ such that 
$\rcx_0\cx{x} \red \exp'$ and $\rcx_1\cx{x} \red \exp'$
where $x$ is a fresh variable, then
$\rcx_0\cx{\exp} \opeq \rcx_1\cx{\exp}$ for any $\exp$.
This rule says that if two reduction contexts have a common 
$\lambda$-reduct when the redex hole is filled with a fresh variable 
(standing for an arbitrary value expression), then they are equivalent.
In other words, two reduction contexts are considered equivalent if placing an arbitrary
expression in the redex hole results in equivalent expressions.

\omitthis{

\begin{verbatim}
  actor computation generated by local actor reductions steps, but
  systems of actors are the star of the show.

  computation in a system has no notion of termination, so
  operational equivalence is must rely on some other notion of
  primitive observable. 

  fairness?

  pattern of messages in and out of a system too complex, need a smaller
  simpler observable.

  moggi laws still hold

  local reduction implies equivalence  (red-exp)

  uniform computation 


\end{verbatim}

In spite of the unpredictability resulting from non-determinism
behavior expressions for actors have a rich equational theory
similar to that developed for Landinesque languages. 

Operational equivalence is defined using the principle of equal observations
in all closing contexts starting with an operational semantics.

An actor configuration is a collection of actors, each with a unique
identifier and associated behavior, together with a set of messages to be delivered.
The configuration is encapsulated by an interface specifying which internal
actors are visible outside, and which external actors are known to the 
configuration.

Several semantics have been defined for actor configurations \cite{talcott-98models}
differing by treatment of ordering relations among send/receive events
\cite{baker-hewitt-77cpp}. The basic operational semantics is the set of traces of fair
executions given by a reduction relation as for Landinesque languages. What is
different is the presence of interactions with the external world -- transitions for input of messages
from external (unseen) actors (\texttt{in(msg)}), and output of messages to these
external actors (\texttt{out(msg)}).   

A closing context for an actor expression is an actor configuration with
empty interface in which
one of the actors is associated with a context in the sense of VTLoE.
  
A special primitive \texttt{event} is added to the language of the observing
context and execution of this primitive is what is observed -- does \texttt{event}
appear in all traces, some traces, or no traces.
Operational equivalence for configurations is defined analogously, where
an observing context is a configuration with a dual interface.

SOMETHING ABOUT REASONING PRINCIPLES -- UNIFORM Computation?

Intuitively actor/interactive computation is more complex/expressive
than sequential computation.  This intuition is tricky to formalize.
In \cite{mason-talcott-05finco} we showed that the trace semantic models of landinesque
expressions are recursively enumerable, while those of actor configurations
are not.

Laws for computational monads hold.
Useful laws for commuting identifying which functional and actor primitives commute, and which actor primitives commute with each other.

}


\section{Cyber-Physical Agents}\label{cp-agents} 

\omitthis{
\begin{verbatim}
Cyber-physical agents
  -- accounting for the physical as well as the cyber
  -- effects on the world
  -- faults and the unpredictable nature of the environment
  -- open challenges 
  -- what are the right models and reasoning principles?
  -- what happens to fairness?
\end{verbatim}
}

Actors are an idealization of real world interactive agents: messages
are always delivered, intact, to the right actor. The interactions
are simply exchanges of information. Autonomous cyber-physical agents
(CPAs) combine interaction as information exchange with interaction
with the physical world via sensors and
actuators. Examples include drones used for agricultural surveillance,
railway track monitoring, or package delivery; security robots, wave gliders that
traverse the Pacific Ocean by themselves; and self-driving cars.  CPAs interact in
space and time and have finite resources. Things don't always work as
expected: sensors may give false readings; actuators (driving
engines, rotors, cameras) may fail to act or cause too much or too
little effect; or there may be natural \emph{threats} such as obstacles
or bad weather impeding a mobile CPA. Communication is likely to be
disrupted so coordination amongst agents is a challenge.

In the actor model the notion of fairness attempts to capture that
actors are independent agents running on independent clocks combined
with reliable message delivery. It ensures that one actor does not get
all the resources in a situation of concurrent processing on a shared
host. In the case of autonomous CPAs we are modeling physically
independent agents. Fairness is in some sense built in to the physics.
Although agents can purposely interfere with one another, that is a
behavior problem, not a model problem.
Also, fairness is an infinitary property, and limits of
the sort used to define fairness aren't observable in the real world.
From a practical point of view, we are typically interested in
behaviors of CPAs over a finite time horizon, in which case fairness,
being an infinitary property, does not play a role.

To define an interaction path semantics for CPA systems, one needs
semantic rules for agent behaviors, which include rules modeling the
physical effects of sensors and actuators, rules modeling relevant
aspects of the external environment.  Examples can be found in
\cite{nigam-etal-17cosim,loreti-hilston-16quanticol}.

To define operational equivalence in analogy to actor systems we would
need a notion of closing configuration. It is not clear that there is
in general a meaningful such notion. If the rules for sensors and
actuators capture fault/threat models they are likely to be
probabilistic, leaving the question of what to check about the set of
interaction paths to decide equivalence. Is it interesting or useful
to have a probability measure on equivalence?

We propose that a first approach to reasoning about CPAs
is to identify effects that we are concerned with,  and use
these to formulate goals 
that a CPA system should achieve. 
Examples of goals include monitoring (taking a picture or sampling air or water for quality assessment, checking inventory); 
moving objects; not running out of energy, not doing damage.
Monitoring goals come with space and time requirements.  
Achievement of goals is not all or nothing, but can be measured
either in a discrete or continuous (partially) ordered domain.
For example the percent of specified locations visited or sampled 
by a monitor system in a given period could be a measure
of achievement.   Another measure could be the percent of energy
remaining or the minimal energy reserve at any point in 
carrying out a task.   These could be combined lexicographically
giving preference to safety to give an overall measure of
success. See~\cite{talcott-arbab-yadav-15wirsing-fest, nigam-etal-17cosim} for examples.  Given such measures, one could compare CPAs based on
how well they achieve goals, leading to a partial order on CPAs.
In different circumstances the ordering of importance of goals may change
and thus the ranking of agents may change.


\section{Conclusion}\label{concl} 

Effects are an essential part of interaction and communication.
In computation systems effects are observed by and affect the
remaining computation (continuation), the concurrent computations,
and observers outside the system.

From studies that develop theories of effects,  key concepts for
formalizing and reasoning about programs with effects have
emerged. These include a variety of contexts
(reduction, memory, closing, \ldots); reduction calculii, and operational
notions of equivalence. The ability to represent execution state
as contexts leads to an elegant operational semantics, and is also
key for further developing the theory of effects.
Uniformity -- reduction rules that are uniformly parameterized by
the surrounding context -- is a powerful tool for developing reasoning
principles; an example of this is the CIU theorem.

Equivalence and the consequences of effects are very sensitive to
the richness of the contexts. Contexts have dimensions beyond
what is normally thought of as effects, including: first-order
versus higher-order, functions can encapsulate and replicate
effects as they are passed around; sequential versus
concurrent/distributed, introducing the complications of
non-determinism and interference mid-computation.

In each case some equational laws will break. However, the laws
of the computational lambda calculus hold in all cases where
there is a uniform semantics, an indication of the importance of
that calculus as a core for computational languages.

Logics for reasoning about programs/systems with effects have
been developed building on the equational theories. Again
contexts are key for axiomatizing effectual primitives and for
expressing properties such as invariants. There are completeness
results for first-order fragments. In other cases a combination
of computational and logical reasoning seems useful, taking
advantage again of reasoning principles based on uniform
computation.

There remain a number of interesting challenges for reasoning
about effects. One example is relating syntactic and semantic
notions of type. Are there semantic types that can be checked by
syntactic type rules? Are there syntactic types that have
semantic characterizations.

Syntactic representation of execution contexts has been a crucial
tool for developing reasoning methods. Although the formal
development has not been done for actor languages, the reasoning
methods relied on a mix of syntactic and semantic contexts that
make it clear a fully syntactic representation of computation
contexts is possible. This remains an open question for
cyber-physical agents (CPAs). Perhaps some form of symbolic
reasoning where the unknown parts of the context remain
unspecified?

Reasoning about CPAs introduces many new issues as a consequence
of the physical nature of effects and interacting in an open
unpredictable environment. Sensors and actuators may be faulty,
other agents and nature may interfere. Furthermore, some
cyber/digital effects disappear when system stops (files,
databases, hopefully do not). Effects caused by CPAs may persist
after the system task ends, by design or due to errors, until
another system (CPA, nature, human) causes further change.

\omitthis{

\begin{verbatim}

Effects are an essential part of interaction and communication.

Observed by 
 continuation
 concurrent agent
 observer out side system 

Cyber/digital effects  disappear when system stops
CP effects may persist after the system task ends,
  until another system acts (CPA, nature, human ...)
  
Logic for reasoning about programs/systems with effects
should express/axiomatize consequences of  executing effect primitives,

The ability to represent execution state as contexts is key.
 Can do for functional
 Almost for actors.
 Open question for CPA

Key concepts
  Contexts
  Reduction rules
  Notions of equivalence  (very sensitive to the richness of the contexts).
  Laws for equivalence -- which effects break what
  Characterizing effects of an expression
  CIU

What are the properties of programs with effects?
What can be captured in a logic?

Actor languages are uniform, but are not Landinesque without further expansions of the language
to construct and manipulate actor configurations.
Same is true CPS only more so. 

Issues/dimensions
  FO vs HO 
  Seq vs Distributed
  Syntactic vs Semantic


\end{verbatim}
}



\bibliographystyle{abbrv}
\bibliography{mt,random}
\end{document}